

\def\a{\alpha}
\def\d{\delta}
\def\e{\epsilon}
\def\l{\lambda}
\def\n{\eta}
\def\t{\tau}
\def\rd{\partial}
\def\L{\Lambda}
\def\S{\Sigma}
\def\A{{\cal A}}
\def\E{{\cal E}}
\def\F{{\cal F}}
\def\G{{\cal G}}
\def\H{{\cal H}}
\def\Hp{{\H_{\perp}}}
\def\P{{\cal P}}
\def\Q{{\cal Q}}
\def\tE{\tilde E}
\def\th{\tilde h}
\def\tp{\tilde p}
\def\hx{\hat x}
\def\hy{\hat y}
\def\pl{\ +\ }
\def\={\ =\ }
\def\and{\ {\rm and}\ }
\def\rtar{\rightarrow}
\def\h{\textstyle {1\over 2}}
\def\Io{1\over}

\magnification\magstep1

\rightline {gr-qc/9405030}
\centerline {Linearized Constraints in the Connection Representation:}
\centerline {  Hamilton-Jacobi Solution.}
\medskip
\centerline {J. N. Goldberg}
\centerline {Department of Physics, Syracuse University, Syracuse, NY
13244-1130}
\centerline {and}
\centerline {D. C. Robinson}
\centerline {Department of Mathematics, King's College, London WC2R 2LS, UK}
\medskip
\centerline {Abstract}
\smallskip
Newman and Rovelli have used singular Hamilton-Jacobi transformations to reduce
the phase space of general relativity in terms of the Ashtekar variables.
Their solution of the gauge constraint cannot be inverted and indeed has no
Minkowski space limit.  Nonetheless, we exhibit an explicit Hamilton-Jacobi
solution of all the linearized constraints.  The result does not encourage
an iterative solution, but it does indicate the origin of the singularity of
the Newman-Rovelli result.
\medskip
\noindent 1.  Introduction. \hfil \break
\medskip
The new variables of Abhay Ashtekar [1,2] for canonical general
relativity, introduced new ways of thinking about the program
to quantize the Einstein theory.  The fact that in terms of
the new variables the constraints are polynomial gives some
hope that it may be possible to find solutions to the
constraints or to implement them as operators on an
appropriate Hilbert space.  Furthermore, because the
configuration space variable is an $SL(2,C)$ connection, it follows
that one can work explicitly with the holonomy operators and
the geometrical concepts related to the loops on which the
holonomy is defined - namely, the knots and linkages of the
loops [3]  .  This has enriched the conceptual arena for the
consideration of quantization of general relativity.

However, new variables do not by themselves eliminate old
problems.  Indeed, the Ashtekar phase space of an $SL(2,C)$
connection, $A^i{}_a$, and a densitized triad, $\tE_i{}^a$,
brings in the Gauss law constraint
$$
D_a\tE_i{}^a \= 0. \eqno (1.1)
$$
in addition to the usual vector and scalar constraints of the
theory.  (Above, the index $i=1-3$ labels the bases for the
$SL(2,C)$ algebra, while $a=1-3$ are indices for coordinates
on a three manifold $\S$.)  In the quantum theory, physical
state vectors are to be independent of the mappings generated
by the constraints.  This means that either the constraints
are to be solved by elimination of some variables, or the
constraints are to be implemented as operators on the physical
Hilbert space.  The important thing is that the state vectors
should only be functionals of those quantities which represent
independent dynamical degrees of freedom of the gravitational
field.

The main thrust of the research so far has been to construct
functionals which are invariant under the constraint mappings.
They depend on the topological structure of loops and not
upon the location or orientation of the loops.  While
operators are defined which act on these functionals, there is
no clear relationship between these operators and the
dynamical degrees of freedom of the gravitational field.
Thus, as yet there is no understanding of how this new view of
general relativity is to be connected to our space-time
picture.

Recently, the Hamilton-Jacobi formalism has been considered as
a way to construct the independent degrees of freedom [4,5].  In
a dynamical system with first class constraints [6-9],
$$
C_r(q_a,\ p^a)\= 0, \qquad r=1\cdots m,\quad a=1\cdots n>m
\eqno (1.2)
$$
the Hamiltonian contains a linear combination of the
constraints:
$$
H = H_0(q_a,\ p^a)\pl \sum_{r=1}^m \l^r\ C_r(q_a,\ p^a) \eqno (1.3)
$$
where the $\l^i$ are arbitrary functions.  As a result, Hamilton's
principal function $S(q_a,P^A,t)$ must satisfy the system of
equations
$$
\eqalignno{
H_0(q_a,\ {{\rd S}\over {\rd q_a}})\pl {{\rd S}\over {\rd t}}
&\= 0 , & (1.4a)\cr
C_r(q_a,\ {{\rd S}\over {\rd q_a}})&\= 0. & (1.4b) \cr}
$$

In a wholly constrained theory,$H_0=0$, there are only the
constraint equations to satisfy.  The resulting canonical
transformation is singular in that there are only $n-m$
constants of integration $P^A$.  Inversion of the
transformation thus depends on $m$ arbitrary functions.  And
in the case of the wholly constrained system, there is no
explicit time dependence.

This formalism is being applied by Newman, Rovelli, and co-
workers [10] to the free electromagnetic field and  the quantum
theory is being worked out in some detail.
Newman and Rovelli have also applied the formalism to
canonical general relativity using the Ashtekar variables [5].
They are able to treat the Gauss law constraints and the
vector constraints, but are unable to invert the canonical
transformations.  Therefore, they are not able to express the
scalar constraint in terms of the thus reduced phase space
variables.

In the following section, we shall sketch the result of Newman
and Rovelli to indicate explicitly  the difficulty with the
inversion and the singularity of the Minkowski space limit.
Then we exhibit the linearized constraints in the Ashtekar
variables and show that we can carry out the Hamilton-Jacobi
transformation for the whole system of constraints.  However,
the method leads, as expected, to non-local variables and
the difficulty of constructing a systematic iteration will
become clear.
\medskip
\noindent 2.  Newman-Rovelli Transformation.
\medskip
In the Ashtekar formalism, the phase space of general
relativity is coordinatized by $(A^i{}_a, \tE_i{}^a)$, an
$SL(2,C)$ connection and a densitized triad on a three-
manifold $\S$. Therefore, there is, in addition to the
diffeomorphism invariance, the local triad invariance.  As a
result we have the Gauss law constraint noted above,
$$
\G_i\ :=\ D_a\tE_i{}^a\= \rd_a \tE_i{}^a \pl \e_{ijk}A^j{}_a \tE_k{}^a \= 0,
\eqno (2.1a)
$$
as well as the vector and scalar constraints:
$$
\eqalignno{
\H_a\ := & F^i{}_ab \tE_i{}^b \= 0, & (2.1b) \cr
\Hp\ := \e_{ijk}& F^i{}_{ab}\tE_j{}^a \tE_k{}^b \= 0, & (2.1b) \cr}
$$
where
$$
F^i{}_{ab} \= 2 A^i{}_{[a,b]}\pl \e^i{}_{jk}A^j{}_a A^k{}_b.
$$

The triad rotations are generated by
$$
G(\L^i)\= \int_{\S}d^3x\ \L^i \G_i
$$
while the mappings of $\S\rtar \S$ are generated by
$$
 H({\vec N})\= \int_{\S} d^3x\ N^a\H_a .
$$
The scalar constraint generates canonical transformations of
the phase space variables which we may identify as evolution.
Therefore, the Hamiltonian is just a linear combination of the
constraints:
$$
H(N,{\vec N},\L^i)\= \int_{\S}d^3x\ \{N \Hp \pl N^a\ \H_a\pl \L^i\G_i\}.
\eqno (2.2)
$$

The singular Hamilton-Jacobi transformations discussed above,
reduce the phase space variables to those which are
coordinates on a constraint surface.  It is possible to do
this step by step, treating each constraint in turn.  That is how Newman and
Rovelli proceed [5].  To treat the Gauss law constraints, one needs variables
which are invariant under the the rotations of the triad field
generated by $\G_i$.  These are obtained from the holonomy of
the connection.  The holonomy group is defined by the parallel
transport around closed loops in $\S$:
$$
U[A,\a]\= P\exp \oint_{\a} A^i{}_a {\dot\a}^a \t_i\ ds, \eqno (2.3)
$$
where the $\t_i$ are the Pauli matrices and $x^a=\a^a(s)$ is
the closed curve.  Clearly, $U[A,\a]$ is gauge covariant,
while its trace, is gauge invariant.  Therefore, one expects
the Hamilton-Jacobi functional to involve the trace.  For the
six new gauge invariant momenta,  they choose six scalar
functions $(u^I,\ v^I),\ I=1-3,$ whose level surfaces
$u^I=\l,\ v^I=\nu$ are assumed to intersect in closed loops,
$\a_I\equiv \a(u^I,v^I)|_{\l,\nu}$.

The Hamilton principal functional is then written as
$$
S[A^i{}_a,u^I,v^I]\= \sum_{I=1}^3\int d\l d\nu\
Tr P\exp\oint_{\a_I} A^i{}_a\t_i{\dot \a}_I{}^a ds^I, \eqno (2.4)
$$
where $s^I$ is the parameter around the loop $\a_I$. It follows that
$$
\tE_i{}^a \= {{\d S}\over {\d A^i{}_a(x)}}\= \sum_I\
\e^{abc}u^I{}_{,b}v^I{}_{,c}\ M_{Ii}(x), \eqno (2.5)
$$
$$
M_{Ii}\ :=\ Tr(U[A,\a_I]\t_i).
$$
and the new configuration variables are defined by (no sum on
I)
$$
\eqalign{
{{\d S}\over {\d u^I(x)}}&\= q_{u^I} \=
\e^{abc}v^I{}{,a}F^i{}_{bc}M_{Ii}, \cr
{{\d S}\over {\d v^I(x)}}&\= q_{v^I} \=
-\e^{abc}u^I{}{,a}F^i{}_{bc}M_{Ii}. \cr} \eqno (2.6)
$$

Note that a general element of $SL(2,C)$ is a linear
combination of the Pauli matrices and the identity.  Thus,
$(a^2_I=1+\sum_{i=1}^3(b_I{}^i)^2),$
$$
U[A,\a_I]\= a_I\ 1\!|\pl b_I{}^i\ \t_i,
$$
so that
$$
M_{Ii}\= 2\ b_I{}^i.
$$
Since in Minkowski space the holonomy group is the identity,
it follows that $M_{Ii}$ is identically zero in the flat space limit.

{}From the equations for $\tE_I{}^a, q_{u^I}, \and q_{v^I}$, it is
evident that the vector constraint can be written in terms of
these gauge reduced variables:
$$
\H_a\= F^i{}_{ab}\tE_i{}^b\= -\h \sum_I(q_{u^I}u^I{}_{,a}\pl
q_{v^I}v^I{}_{,a}). \eqno (2.7)
$$

To treat the vector constraint, one needs to construct 3-
diffeomorphism invariant quantities.  One way to do so is to
introduce, on the manifold $\S$, intrinsic coordinates which
are defined by an algorithm.  In terms of such coordinates,
the points of the manifold are uniquely labled and other
scalar functions of the intrinsic coordinates are then truly
invariant.  In this case there are six scalars.  Choose the
$v^I$ to be the intrinsic coordinates and then
$$
U^I(v^J(x))\= u^I(x) \eqno (2.8)
$$
will be the sought for invariants.  The Hamilton-Jacobi equations are
$$
\sum_I\ (u^I{}_{,a}{{\d S}\over {\d u^I}}\pl
v^I{}_{,a}{{\d S}\over {\d v^I}})\= 0
$$
which has the solution
$$ S(u^I,v^I,P_J)\= \sum_I\int_{\S}d^3x\ u^I(x)\ P^I(v^J(x))
|{{\d v}\over {\d x}}|.  \eqno (2.9)
$$
{}From this functional we find
$$
\eqalign {q_{u^I}(x)&\= {{\d S}\over {\d u^I(x)}}\=
|{{\d v}\over {\d x}}|P_I(v^J(x)), \cr
q_{v^I}(x)&\= {{\d S}\over {\d v^I}}\= -|{{\d v}\over
{\d x}}|\sum_K\ P_K(v^J(x))u^K{}_{,a}{{\rd x^a}\over {\rd
v^I}} . \cr} \eqno (2.10)
$$
and for the new coordinates
$$
Q^I(v)\= {{\d S}\over {\d P_I}}\= u^I(x)\= U^I(v(x)). \eqno (2.11)
$$

The $(P_I(v), Q^I(v))$ are the coordinates on the gauge and
3-diffeomorphism reduced phase space.  One would like to treat
the scalar constraint in a similar fashion.  However, we don't
know how to write the scalar constraint in terms of these
reduced coordinates.  It must be possible to do so, but it
appears to require the inversion of the canonical
transformations.  The problem goes back to the Gauss law
transformation.  There we see that the relationships among the
old and new variables involves the triad, the connection, and
some information about the holonomy of the loops $\a_I$.
But, this appears in such a convoluted fashion that it does
not even allow one to see how to construct even an implicit
solution.  Some deep insight is needed at this point.

We noted earlier that the flat space limit does not exist and is,
in fact, singular.  In flat space the triad is not zero, but
constant.  On the other hand, the right hand side of Eq (2.5) is
zero.  Thus, even a solution by iteration is precluded.
However, in the next section, we shall find a solution for the
linearized constraints directly and not as a limit of the
Newman-Rovelli result.
\medskip
\noindent 3.  Linearized Constraints.
\medskip
The Gauss law constraint written out reads
$$
\tE_i{}^a{}_{,a} \pl \e_{ijk}A^j{}_a \tE_k{}^a\= 0.  \eqno (3.1)
$$
For the Minkowski space solution we choose
$$
{}^0\!\tE_i{}^a=\d_i{}^a \and {}^0\!A^j{}_a=0. \eqno (3.2)
$$
Therefore, we write
$$
\tE_i{}^a\= \d_i{}^a\pl \E_i{}^a, \qquad A^i{}_a\= \A^i{}_a, \eqno (3.3)
$$
where the calligraphic letter is the difference of the quantity
from
its Minkowski space value.  Therefore, we have
$$ \eqalignno{
\E_i{}^a{}_{,a} \pl \e_{ijk}\A^j{}_a\d_k{}^a &\= 0, &(3.4a)\cr
\F^i{}_{ab}\d_i{}^a &\= 0, & (3.4b) \cr
\e_{ijk}\F^i{}_{ab}\d_j{}^a\d_k{}^b&\= 0. &  (3.4c) \cr
}$$
for the Gauss law, the vector constraint, and for the scalar constraint,
respectively.  In the above, we have used the linearization
$$
F^i{}_{ab} \= \F^i{}_{ab} \= \A^i{}_{b,a}-\A^i{}_{a,b}.
$$

Although there is no need to write a transformation for the zeroth
order Gauss law equation, it is convenient to express the solution
in terms of the scalars ${}^0\!u^I \and {}^0\!v^I$.  Thus, mod 3, we have
$$
{}^0\!u^i\= x^{i+1},\qquad {}^0\!v^i\= x^{i+2}
$$
so that $ {}^0\!E_i{}^a\= \e^{abc} {}^0\!u^i{}_{,b} {}^0\!v^i_{,c}$.  We have
used the fact that in the linearization, the index $I$ can be replaced by the
group index $i$.  In the following, the summation convention will not apply to
the group indices $i,j,\cdots$. The summation will be explicitly exhibited.

In Eq. (3.5), we give the solution for the linearized Gauss law as the general
solution of the homogeneous equation and a particular solution of the
inhomogeneous equation.  It is useful to write out that solution
extensively as follows ($u\and v$ below refer to linearized
values):
$$
\eqalign{
\E_1{}^a&\= \e^{abc} (u^1{}_{,b}\d^3_c \pl \d^2_b v^1{}_{,c}) \pl \cr
&\qquad\h\Bigl\{
\d_2{}^a\{\int_{-\infty}^{x^2}- \int_{x^2}^{\infty}\}
\A^3{}_2 dx^{2'}\- \d_3{}^a\{\int_{-\infty}^{x^3}-
\int_{x^3}^{\infty}\}\ \A^2{}_3 dx^{3'} \Bigr\}, \cr
\E_2{}^a&\= \e^{abc} (u^2{}_{,b} \d^1_c \pl \d^3_b v^2{}_{,c}) \pl \cr
&\qquad\h\Bigl\{
\d_3{}^a\{
\int_{-\infty}^{x^3}- \int_{x^3}^{\infty}\}
\A^1{}_3 dx^{3'}\- \d_1{}^a\{\int_{-\infty}^{x^1}-
\int_{x^1}^{\infty}
\}\ \A^3{}_1{} dx^{1'} \Bigr\}, \cr
\E_3{}^a&\= \e^{abc} (u^3{}_{,b} \d^2_c \pl \d^1_b v^3{}_{,c}) \pl \cr
&\qquad\h\Bigl\{
\d_1{}^a\{\int_{-\infty}^{x^1}- \int_{x^1}^{\infty}\}
\A^2{}_1 dx^{1'}\- \d_2{}^a\{
\int_{-\infty}^{x^2}- \int_{x^2}^{\infty}
\}\ \A^1{}_2 dx^{2'} \Bigr\}.   \cr} \eqno (3.5)
 $$

{}From the above, it is easy to show that the following Hamilton-Jacobi
functional
$$\eqalign{
S(\A^i{}a, u^i, v^i)&\= \sum_i\int_{\S} d^3x \e^{abc}\bigl(
u^i{}_{,b}\d^{I+2}_c- \d^{i+1}_b v^i{}{,c}\bigr) \A^i{}_a \pl \cr
&\qquad{\Io 4}\Bigl\{\int_{\S} d^3x\A^1{}_a(x)\Bigl[\d^a{}_2\bigl(
\int_{-\infty}^{x^2} - \int_{x^2}^{\infty}\bigr) dx^{2'}\
\A^3{}_2(x^1,x^{2'},x^3) \- \cr
&\qquad\d^a{}_3\bigl(\int_{-\infty}^{x^3} -
\int_{x^3}^{\infty}\bigr) dx^{3'}\ \A^2{}_3(x^1,x^2,x^{3'})\Bigr]
\pl  \cr
&\qquad\int_{\S} d^3x\A^2{}_a(x)\Bigl[
\d^a{}_3\bigl(\int_{-\infty}^{x^3} - \int_{x^3}^{\infty}\bigr)
dx^{3'}\ \A^1{}_3(x^1,x^2,x^{3'}) \- \cr
&\qquad\d^a{}_1\bigl(\int_{-\infty}^{x^1} -
\int_{x^1}^{\infty}\bigr) dx^{1'}\ \A^3{}_1(x^{1'},x^2,x^3)\Bigr]
\pl \cr
&\qquad\int_{\S} d^3x\A^3{}_a(x)\Bigl[\d^a{}_1\bigl(
\int_{-\infty}^{x^1} - \int_{x^1}^{\infty}\bigr) dx^{1'}\
\A^2{}_1(x^{1'},x^2,x^3) \- \cr
&\d^a{}_2\bigl(\int_{-\infty}^{x^2} - \int_{x^2}^{\infty}\bigr)
dx^{2'}\ \A^1{}_2(x^1,x^{2'},x^)\Bigr] \Bigr\}. \cr} \eqno (3.6)
$$
generates a canonical transformation which gives the above
solution for $\E_i{}^a$ and
satisfies the Gauss law constraint,
$$
{\rd \over {\rd x^a}} \Bigl\{{{\d S}\over {\d \A^i{}_a}}\Bigr\}
\pl \sum_{i,j=1}^3\e_{ijk}\A^i{}_a\d_k{}^a \= 0.
$$

We now obtain the new configuration space variables from
$S(\A^i{}_a,u^i, v^i)$,
$$\eqalignno{
{{\d S}\over {\d u^i}}&\= q_{u^i}\= B_i{}^a{}^0\!v^i{}_{,a} \=
B_i{}^{i+2}, &(3.7a) \cr
{{\d S}\over {\d v^i}}&\= q_{v^i}\= -B_i{}^a{}^0\!u^i{}_{,a} \=
-B_i{}^{i+1}.    &(3.7b)\cr
}$$
In the above, $B^{ia}\= \h \e^{abc}\F^i{}_{bc}$.

Note that the linearized vector and scalar constraints are homogeneous in
$B^{ia}$, so that only the zeroth order part of $E_i{}^a$ appears in Eqs.
(3.7).  As a result,
the inhomogeneous part of $\E_i{}^a$ does not appear in the remaining canonical
transformations.  This result makes it easier to carry out the remaining
transformations.

It is easy to see that the vector constraints take the simple form
$$
\H_a\= \sum_i\ (q_{u^i}\d^i{}_{i+1} \pl q_{v^i}\d^i{}_{i+2})\=0  \eqno (3.8)
$$
and the scalar constraint
$$
\Hp\= \sum_i B^i{}_i \= 0. \eqno (3.9)
$$
Note that
$$
P^i\ :=\ v^{i+1}\- u^{i+2}  \eqno (3.10)
$$
has a vanishing Poisson bracket with $\H_a$ so that the Hamilton-Jacobi
function
$$
S(u^i,v^i,Q_i)\= -\int_{\S} d^3x \sum_i Q_i(v^{i+1}\- u^{i+2}), \eqno (3.11)
$$
which leads to the transformation
$$
q_{u^i}\= -Q_{i+1} \qquad \and \qquad q_{v^i}\= Q_{i+2},  \eqno (3.12)
$$
clearly satisfies the vector constraint.  The new momenta are precisely
the expressions given in Eq. (3.10).

{}From Eqs. (3.7)  above, we see that the components of $B^i{}^a$ for
$a\ne i$ are given in terms of $q_{u^i} \and q_{v^i}$, hence
in terms of the reduced phase
space variables $(Q_i, P^i)$.  Now, from its definition, $B_i{}^a{}_{,a}=0$.
Therefore, we can express $B_i{}^a$ in terms of a potential $\Q_i{}^{ab}$
which is antisymmetric in the indices $(a,b)$.  Furthermore, we find that
this potential vanishes unless either $a=i$ or $b=i$ so that the non-zero
components are given by (no sum on $i$)
$$
\Q_i{}^{ib}\= \int_{-\infty}^{x^i} (Q_{i+2}\d^a{}_{i+1}
\pl Q_{i+1}\d^a{}_{i+2}) dx^{i'}.  \eqno (3.13)
$$
The scalar constraint is now given by
$$
\Hp\=\Q^a{}{,a}\= 0,\qquad {\hbox{\rm where}}\qquad
\Q^a\ :=\ \sum_i\ \Q_i{}^{ia}. \eqno (3.14)
$$
To find the final singular transformation which leads to the fully reduced
phase space, it is convenient to carry out a non-singular canonical
transformation which makes the $\Q^a$ the new canonical coordinates.  For this
purpose, consider
$$
G(Q_i, \P_a)\= \int_{\S}d^3x \Q^a(Q_i) \P_a,
$$
which leads to $\Q^a$ as the new coordinates and
$$
\P_i\= {1\over 2}\{\rd_{i+2}P_{i+1}\pl \rd_{i+1}P_{i+2} \-
\int_{-\infty}^{x^i} \rd_{i+1}\rd_{i+2} P_i dx^{i'} \}. \eqno (3.15)
$$
where we have used $\rd_i:=\rd/\rd x^i$.
We can now write the Hamilton-Jacobi functional for the scalar constraint in
terms of the momenta $\P_a$ and two functions $U \and V$ and two constant
one forms $\mu_a \and \nu_a$,
$$
S(\P_a,U,V) \= -\int d^3x \e^{abc}(\mu_cU_{,b} \pl \nu_bV_{,c} )\P_a
\eqno (3.16)
$$
so that
$$
\Q^a \= \e^{abc}(\mu_cU_{,b} \pl \nu_bV_{,c})  \eqno (3.17)
$$
identically satisfies the remaining constraint.  The new momenta are given by
$$
\Pi_U \= \e^{abc}\mu_c\P_{a,b}\qquad \Pi_V \= -\e^{abc}\nu_c\P_{a,b}.
\eqno (3.18)
$$

The totally reduced phase space is described by the four functions $U, V,
\Pi_U \break \and \Pi_V$.  Since the Hamiltonian is now zero, there are no
dynamical
restrictions on these functions.  They are to be defined as the quantum
operators in the quantization of this linearized theory.
\medskip
\noindent 4.  Discussion.

We have applied Hamilton-Jacobi theory to the linearized constraints of general
relativity in terms of the Ashtekar variables.  Unlike the treatment by Newman
and Rovelli [5] of the exact formalism which has no flat space limit, we are
able to show that we can use the singular Hamilton-Jacobi transformations to
eliminate the constraints.  In this way we arrive at a set of conjugate
variables for the fully reduced phase space of the linear Einstein equations.
If one could do the same for the exact theory, the inversion of these
transformations would give the general solution of the Einstein equations.
That is not the case for the linearized theory or any truncated version.  The
reason is that the linearized equations for general relativity are obtained
from a quadratic Hamiltonian to which one must adjoin the linearized
constraints with Lagrange multipliers.  Thus, the linearized theory is not a
wholly constrained theory.  The Hamiltonian is not zero to first order.  As a
result, the reduced phase space variables are only a ``gauge invariant'' set of
initial conditions.  The quadratic Hamiltonian then generates the evolution of
these variables.  However, it is a ``gauge invariant'' set which may become the
operators in a quantum theory.

One can see this clearly if one applies this technique to the ADM [11]
formalism.
Remember, the phase space variables are defined on a three-surface, not in
space-time.  So the linearized phase space variables are
$$
h_{ab}\= q_{ab}\- \n_{ab}\qquad \and p^{ab}. \eqno (4.1)
$$
where $q_{ab}$ is the metric on the three-surface $\S$ and $p^{ab}$ is
esentially the linearized part of the extrinsic curvature of $\S$ when embedded
in the space-time defined by the Einstein equations.  The linearized
constraints become
$$
\eqalign{
\H_a&\= -2p_a{}^b{}_b\=0, \cr
\Hp &\= (\n^{ab}\n^{cd}-\n^{ac}\n^{bd})h_{ab,cd} \= 0. \cr}
\eqno (4.2)
$$
If one carries out a three-dimensional Fourier transform of $h_{ab}\and
p^{ab}$, the Hamil\-ton-Jacobi treatment leads to the result that the reduced
phase space is defined by $(h_+, h_{\times}, p^+,p^{\times})$ so that
$$
\eqalign{
\th_{ab}({\vec k})&\= h_+(\hx\hx-\hy\hy)\pl h_{\times}(\hx\hy+\hy\hx) \cr
\tp^{ab}({\vec k})&\= p^+(\hx\hx-\hy\hy)\pl p^{\times}(\hx\hy+\hy\hx) \cr}
\eqno (4.3)
$$
where $\th^{ab} \and \tp^{ab}$ are the Fourier coefficients and ${\vec k}\cdot
\hx={\vec k}\cdot \hy = 0$.  Of course, this is the expected result, but
nothing
tells us that $\vec k$ is the spatial part of a null vector.  That information
comes from the linearized propagation equations.

The result we have suggests that for space-times which have a flat space limit,
we again write $\tE_i{}^a=\d_i^a+\E_i{}^a$ so that the Gauss law constraints
have the form
$$
\rd_a\E_i{}^a\- \e_{ijk}A^j{}_a\E_k{}^a \= \e_{ijk}A^j{}_a\d_k^a. \eqno (4.4)
$$
The Newman-Rovelli solution solves the homogeneous equation, but the
inhomogeneous equation can be solved with the help of line integrals using the
parallel propagator.  To proceed in this manner would lead to much more
complicated expressions for the vector and scalar constraints for which we
doubt that solutions can be found.  But, this does exhibit in part the source
of the singularity of the Newman-Rovelli solution.

It is clear that we were able to complete the transition to the reduced phase
space for two reasons.  One is that we only considered the linear terms of the
constraints.  But, the other is certainly connected with the fact that we did
not need the general solution of the Einstein equations.  The exact theory is
wholly constrained so that solving the singular Hamilton-Jacobi problem is
equivalent to finding the general solution of the Einstein equations.  In
principle, the work by Kozameh and Newman [12] does give this general solution
in
terms of data on future null infinity.  However, it is not clear how to make
use of this result.
\medskip
\noindent Acknowledgments.

The work of JNG was supported in part by the NSF under Grant PHY 9005790.
Both authors wish to thank SERC for support of JNG by a SERC Visiting
Fellowship Grant GR/H56472 during the summer and fall term of 1993.  JNG is
grateful for the hospitality of the Department of Mathematics at King's College
where this work was begun.
\medskip
\noindent References.

\item {1.}  A. Ashtekar, Phys. Rev. Lett. {\bf 57}, 2244 (1986).
\item {2.}  A. Ashtekar, {\it Lectures on Non-perturbative Canonical Gravity
(Notes prepared in collaboration with R. Tate)},(World Scientific, Singapore,
1991).
\item {3.}  C. Rovelli and L. Smolin, Nucl. Phys. {\bf B331}, 80 (1990).
\item {4.}  J. N. Goldberg, E. T. Newman, and C. Rovelli, J. Math. Phys.
{\bf 32}, 2739 (1991).
\item {5.}  E. T. Newman and C. Rovelli, Phys. Rev. Lett. {\bf 69}, 1300
(1992).
\item {6.}  P. G. Bergmann, Phys. Rev. {\bf 144}, 1078 (1966).
\item {7.}  A. Komar, Phys. Rev. {\bf 170}, 1175 (1968).
\item {8.}  A. Komar, Phys. Rev. {\bf D1}, 1521 (1970).
\item {9.}  C. Teitelboim and M. Henneaux, {\it Quantization of Gauge Systems}
(Princeton University Press, Princeton, N. J., 1992).
\item {10.}  S. Frittelli, S. Koshti, E. T. Newman, and C. Rovelli, ``Classical
and Quantum Dynamics of the Faraday Lines of Force'', (University of Pittsburgh
preprint, December, 1993).
\item {11.}  R. Arnowitt, S. Deser, and C. Misner, in {\it Gravitation}, ed. L.
Witten (John Wiley \& Sons, New York, 1962).
\item {12.}  C. N. Kozameh, E. T. Newman, and S. V. Iyer, J. Geom. Phys.
{\bf 8}, 195 (1992).
\end